# Physics-constrained indirect supervised learning

**Yuntian Chen**, Intelligent Energy Lab, Frontier Research Center, Peng Cheng Laboratory, Shenzhen 518000, China; **Dongxiao Zhang**, School of Environmental Science and Engineering, Southern University of Science and Technology, Shenzhen 518055, China

*Abstract*: This study proposes a supervised learning method that does not rely on labels. We use variables associated with the label as indirect labels, and construct an indirect physics-constrained loss based on the physical mechanism to train the model. In the training process, the model prediction is mapped to the space of value that conforms to the physical mechanism through the projection matrix, and then the model is trained based on the indirect labels. The final prediction result of the model conforms to the physical mechanism between indirect label and label, and also meets the constraints of the indirect label. The present study also develops projection matrix normalization and prediction covariance analysis to ensure that the model can be fully trained. Finally, the effect of the physics-constrained indirect supervised learning is verified based on a well log generation problem.

## I. Introduction

In machine learning, supervised learning often needs to directly obtain the specific output value (label) of each data[1,2]. However, the label is sometimes difficult to obtain in practice, and the dependence on labeling has restricted the application of supervised learning in practical engineering problems.

The essence of the algorithms and models is to extract information from the training data, and the supervised learning indicates that the information is extracted from the data with labels. When training a model, a specific label is not necessarily required as the output of the model[3]. In fact, only information describing the distribution of the output is needed. This is similar to the fact that humans are able to learn without direct labels, and only need a description of what an output should be[3]. Conventional supervised learning is a completely data-driven method, in which the labels can simply and directly describe the distribution of the output. However, the information used to train the model can not only come from the data itself, but also from *a priori* information, such as physical mechanisms[4-7]. In this study, the motivation of indirect supervised learning is to combine the information of physical mechanisms and data. We show, in the following sections, that it is possible to perform supervised learning based on indirect variables and prior information without relying on specific labels.

Specifically, it can be assumed that the input variables of the model are represented by **X**, the output variables to be predicted are **Y**, and a mapping relationship exists between **X** and **Y** defined by function $f(\cdot)$. In conventional supervised learning, the model is directly trained with the data distribution of both **X** and **Y**. Nevertheless, if we can find variables **H**, and there is a mapping $g(\cdot)$ between **H** and **Y**, then we can also attempt to train a model that predicts **Y** based on the data distribution of **X** and **H** combined with the information of the mapping function $g(\cdot)$ that holds over the output space. In other words, **H** is the indirect label used by the model. The information of $g(\cdot)$ is utilized to ensure that the output of the neural network conforms to the mapping relationship between **H** and **Y**, and then **Y** is substituted by **H** to train the model. The relationship between **X**, **Y**, and **H** is shown in Figure 1.

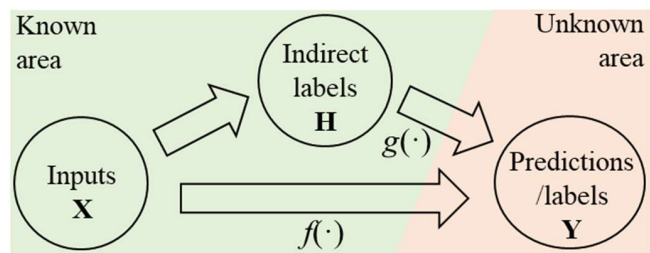

**Fig. 1.** Relationship between inputs, labels, and indirect labels.

In order to introduce indirect supervised learning more intuitively, the present study takes the problem of synthetic well log generation as an example. Previous studies have indicated that the neural network can generate well logs[8-11]. Zhang et al. utilized an LSTM network which is capable of processing sequential data, and the model performance is better than that of conventional neural networks[12]. Here, we predict the uniaxial compressive strength (UCS) directly based on conventional well logs, including depth, density, resistivity, and gamma rays via LSTM. Because UCS does not exist in the dataset, we cannot directly use UCS as a label to construct a training dataset for supervised learning. However, if the sonic logs are available, we can employ them as indirect labels since a mapping relationship exists between the sonic logs and the UCS. Therefore, the sonic logs can be used as indirect labels and the conventional logs can be taken as inputs to directly train a model to predict the UCS based on physical mechanisms. It should be mentioned that, although the aforementioned method belongs to supervised learning, it does not utilize the label (UCS) directly, but it is based on indirect labels, and thus this method is called physics-constrained indirect supervised learning.

## II. METHODOLOGY

### A. Projection matrix based on physical mechanisms

Indirect labels cannot be used directly to train the model. The mapping relationship between sonic logs and the UCS can be split into non-linear and linear parts according to the physical mechanism. Firstly, the dynamic Young's modulus ($E_{dyn}$) can be obtained by Eq. 1 based on sonic logs (indirect label) and density log (input), where $c$ represents a constant, $\rho$ is the density, the $\Delta t_s$ and $\Delta t_p$ are the sonic logs[13]:

$$E_{dyn} = c \cdot \left(\frac{\rho}{\Delta t_s^2}\right) \cdot \left(\frac{3\Delta t_s^2 - 4\Delta t_p^2}{\Delta t_s^2 - \Delta t_p^2}\right) \quad (1)$$

When calculating the UCS, the static Young's modulus $E_{stat}$ is usually used instead of the $E_{dyn}$. The $E_{stat}$ is obtained through experiments, and it can be estimated by $E_{dyn}$, as shown in Eq. 2[13]. Finally, the UCS can be calculated by Eq. 3[14]:

$$E_{stat} = 0.414 \cdot E_{dyn} - 1.05 \quad (2)$$
$$UCS = 2.28 + 4.1089 \cdot E_{stat} \quad (3)$$

In this study, the model is trained through physical mechanisms based on indirect labels. Therefore, based on Eq. 1, Eq. 2, and Eq. 3, the problem can be expressed in the form of matrixes, as shown in Eq. 4, where $\mathbf{A} \in \mathbf{R}^{n \times 2}$ is constructed based on the dynamic Young's modulus, $\mathbf{Y}$ represents the UCS, $\mathbf{X}$ represents the inputs, and $n$ is the number of training data (the product of batch size and training data length):

$$\mathbf{Y} = \mathbf{AX} \quad \text{where } \mathbf{A} = \begin{bmatrix} E_{dyn,1} & \cdots & E_{dyn,n} \\ 1 & \cdots & 1 \end{bmatrix}^T \quad (4)$$

Because $n$ is usually greater than 2 in practice, the matrix $\mathbf{A}$ is strictly skinny, and Eq. 4 is overdetermined, which means that it is difficult to find a set of solutions that make the model fit all of the data points. In other words, there is no guarantee that the predictions are all accurate, and there should be non-zero residuals. Therefore, we define the residual $\mathbf{r} = \mathbf{AX} - \mathbf{Y}$ and attempt to approximately solve $\mathbf{Y} = \mathbf{AX}$ by finding an $\mathbf{X}^*$ that minimizes the residual $\mathbf{r}$. To minimize the residual, we calculate the gradient of the squared residual w.r.t. $\mathbf{X}$, which is shown in Eq. 5:

$$\nabla_x \mathbf{r}^2 = \nabla_x \left(\mathbf{X}^T \mathbf{A}^T \mathbf{AX} - 2\mathbf{Y}^T \mathbf{AX} + \mathbf{Y}^T \mathbf{Y}\right) = 2\mathbf{A}^T \mathbf{AX} - 2\mathbf{A}^T \mathbf{Y} \quad (5)$$

It is known that when the residual is at its minimum, the gradient of the residual must be zero. Since the matrix $\mathbf{A}^T \mathbf{A}$ is usually invertible in practice, the $\mathbf{X}^*$ can be obtained as $\mathbf{X}^* = (\mathbf{A}^T \mathbf{A})^{-1} \mathbf{A}^T \mathbf{Y}$. Based on $\mathbf{X}^*$ and Eq. 4, the relationship between a given prediction value $\mathbf{Y}$ and its projection $\mathbf{Y}^*$ on range($\mathbf{A}$) can be constructed, as shown in Eq. 6. The matrix $\mathbf{P} = \mathbf{A}(\mathbf{A}^T \mathbf{A})^{-1} \mathbf{A}^T$ is called the projection matrix[15], which is the key to indirect label supervised learning. Through Eq. 6, the predicted value can be projected to a numerical space (range($\mathbf{A}$)) that conforms to a certain physical mechanism:

$$\mathbf{Y}^* = \mathbf{AX}^* = \mathbf{A}\left(\mathbf{A}^T \mathbf{A}\right)^{-1} \mathbf{A}^T \mathbf{Y} = \mathbf{PY} \quad (6)$$

Eq. 6 and projection matrix $\mathbf{P}$ have clear physical meanings. Minimizing the residual is equivalent to finding the projection of $\mathbf{Y}$ on range($\mathbf{A}$), i.e., the point where $\mathbf{Y}$ is closest to range($\mathbf{A}$), as shown in Figure 2. Range($\mathbf{A}$) is actually the solution space determined by Eq. 1, Eq. 2, and Eq. 3. It is a set of points that conform to the physical mechanism and is shown as a red line. If the prediction result $f(x)$ of the neural network is taken as $\mathbf{Y}$ and substituted into Eq. 6, then $\mathbf{Y}^*$ is the point in range($\mathbf{A}$) closest to the prediction result of the neural network. Therefore, in essence, the physical meaning of Eq. 6 is to find the point closest to the predicted value of the neural network among all points that meet given physical constraints. The physical constraints used in this process are determined based on prior knowledge and indirect labels.

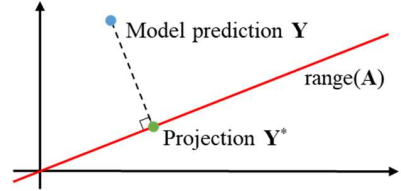

**Fig. 2.** Geometric interpretation of the projection on range($\mathbf{A}$).

### B. Indirect physics-constrained loss

In order to make the prediction result of the neural network comply with physical constraints, a new loss function is constructed based on Eq. 6. The proposed loss function trains the neural network through physical constraints based on known indirect labels, as shown in Eq. 7. This loss is called indirect physics-constrained loss, and it is utilized to calculate the difference between the projected result $\mathbf{Y}^*$ of the predicted value of the neural network in the physics-constrained space and the predicted value of the neural network $\mathbf{Y}$. By iteratively optimizing the neural network through the proposed indirect physics-constrained loss function, the final prediction result of the network can be guaranteed to meet the given physical constraints (i.e., the difference between the projected value $\mathbf{Y}^*$ and the predicted value $\mathbf{Y}$ is the smallest). In addition, the indirect physics-constrained loss is a function of the predicted value of the neural network and does not require the value of UCS, which reflects the meaning of indirect label supervision.

$$Loss = g(f(x)) = \sum \left(\mathbf{Y}^* - f(x)\right)^2 = \sum \left(\mathbf{P}f(x) - f(x)\right)^2 \quad (7)$$

In practical application of the indirect physics-constrained loss, two challenges exist that may cause neural network training to fail. The first challenge is that the projection matrix will change the mean and variance of the distribution of the predictions, and the second challenge is the symmetry of the indirect physics-constrained loss. The above two challenges can be solved by projection matrix normalization and prediction covariance analysis, respectively.

In the first challenge, although the distribution of the projected value $\mathbf{Y}^*$ will meet the physical constraints (range($\mathbf{A}$)), the projection matrix cannot guarantee that the projected value $\mathbf{Y}^*$ and the model prediction $\mathbf{Y}$ are at the same scale. In other words, the $\mathbf{Y}^*$ obtained by the projection matrix conforms to physical constraints in the trend, but its mean and variance differ from those of neural network prediction $\mathbf{Y}$. In practice, the $\mathbf{Y}^*$ calculated based on the projection matrix is often smaller than $\mathbf{Y}$, which means that the loss function tends to reduce the scale of the model output. This will cause the predicted value of the neural network to gradually decrease as the iteration progresses, eventually approaching 0. The small model prediction value in the loss function, in turn, will cause the loss value to be extremely small, making the loss function unable to effectively provide the optimization direction and result in training failure.

The idea to solve this problem is similar to that of the batch normalization layer, i.e., to perform a normalization after each iteration. This operation can ensure that both the projected value and the predicted value are within a reasonable range, which is conducive to accurately calculating the loss of the model prediction. Specifically, in each iteration, the $\mathbf{Y}^*$ that conforms to the physical mechanism after projection is first normalized, and then the difference with the predicted value of the neural network is calculated and used as the indirect physics-constrained loss. Therefore, normalizing the projection result is conducive to model convergence, and can solve the first challenge.

The second challenge is the symmetry of the indirect physics-constrained loss. This symmetry comprises two parts. On the one hand, the projection matrix based on the indirect label ($E_{dyn}$) is symmetrical to the indirect label, i.e., $\mathbf{P}(E_{dyn}) \equiv \mathbf{P}(-E_{dyn})$. This means that when the indirect label is flipped along the x-axis, the projection result of the neural network prediction does not change. In other words, the model obtained based on the indirect physics-constrained loss has symmetric equivalent solutions. The second symmetry is that the indirect physics-constrained loss is symmetrical to the model predicted value, i.e., the indirect physics-constrained loss $g(f(x)) \equiv g(-f(x))$. This is because all terms in the indirect physics-constrained loss are calculated based on $f(x)$. Consequently, when $f(x)$ takes the opposite number, the projection result will also change, so that the root of the entire loss becomes the opposite number, while the loss itself remains unchanged. Because the indirect physics-constrained loss exhibits the above two symmetries, if the neural network is trained directly by the indirect physics-constrained loss (Eq. 7), there will be a 50% probability that an axisymmetric solution to the trend of the real result will be obtained, as shown in Figure 3. To avoid the effects of randomness, 50 independent repeated experiments were performed. The black curve is the real result, and the red curves and the blue curves are the final results obtained in independent repeated experiments. The results show that both solutions occur exactly 25 times, which indicates that the probability of the two solutions is approximately 50%. The experimental results are consistent with the theoretical analysis.

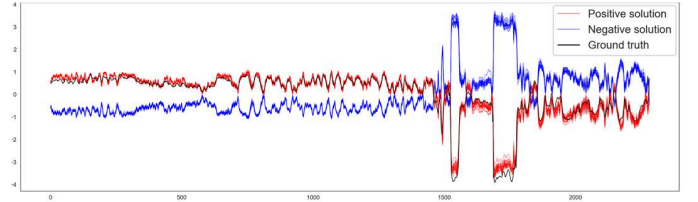

**Fig. 3.** The symmetric equivalent solutions obtained by indirect physics-constrained loss of 50 independent repeated experiments.

This challenge can be solved through covariance analysis. The motivation of this method is that, although the symmetry of indirect physics-constrained loss results in two symmetrical equivalent solutions, the correlation between the true solution and the indirect labels should be the same as the correlation between the variable to be predicted and the indirect label. Specifically, since we know the value of the indirect labels (sonic logs) and the mapping relationship $g(\cdot)$ between the indirect label and the parameter to be predicted (UCS) according to physical mechanism, we can know whether the indirect labels are positively or negatively related to the variable to be predicted. At the same time, the covariance between the model prediction result and the indirect labels can be directly calculated. Therefore, according to the positive and negative covariance between the model prediction result and the indirect labels, the two equivalent solutions can be divided into positive and negative solutions. Specifically, if the variable to be predicted is positively correlated with the indirect label, the positive solution is the true solution; otherwise, the negative solution is the true solution. Through covariance analysis, the true solution in the equivalent solutions of the model can be quickly determined, and the multiple solution problem caused by indirect physics-constrained loss is solved.

### III. EXPERIMENTS

In order to verify the feasibility of the physics-constrained indirect supervised learning, a synthetic well log generation experiment was performed in this study. The experimental data come from 39 wells in North Dakota in the United States, including the true vertical depth log, density log, resistivity log, gamma ray log, and sonic logs[16]. The UCS, which is a kind of geomechanical log, is predicted, but it does not exist in the data set. The depth, density, resistivity, and gamma ray are used as inputs in the model. The indirect labels are sonic logs (Delta-T shear, $\Delta t_s$ and Delta-T compressional, $\Delta t_p$). The entire model does not require UCS values. Instead, it employs the mapping relationship between sonic logs (indirect labels) and UCS (label) as prior information, and uses the indirect physics-constrained loss to train neural networks.

In the physics-constrained indirect supervised learning, projection matrix normalization and prediction covariance analysis are used to ensure the convergence of the model. In this study, the LSTM is utilized as the prediction model to generate synthetic well logs, since it possesses the advantage of processing long-term dependent sequential data. The data set contains a total of 39 wells. Each experiment randomly selects 27 wells (70%) as the training data set and 12 wells (30%) as the test data set. Each experiment was repeated 10 times to avoid the impact of randomness. The hyperparameters involved in the experiment include batch size, training data length, and whether or not to add a batch normalization layer.

In general, a larger batch size is helpful for each batch to better represent the distribution of the whole data set, but a batch size that is too large will also lead to a lack of randomness in the training process, and it is easier to fall into a local optimum. For training data length, this is mainly related to the stratum. Overall, the longer is the training data length, the greater is the stratum thickness corresponding to each training data, but the higher is the computational complexity. The projection matrix in the indirect label supervised learning involves calculating the inverse matrix of $\mathbf{A}^T\mathbf{A}$, and the number of columns is the product of the batch size and the training data length. Therefore, an excessively large batch size and training data length may cause insufficient GPU memory.

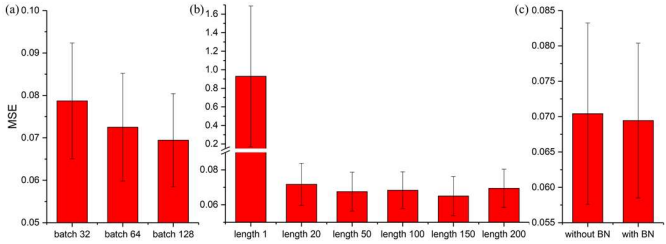
**Fig. 4.** Histogram of MSE with different hyperparameters.

We optimized the hyperparameters of the model separately. The default setting of the hyperparameters is 128 for the batch size, and 200 for the training data length with the batch normalization layer. Firstly, the effects of batch sizes of 32, 64, and 128 are evaluated, respectively, and the results are shown in Figure 4a. The height of the bars in the histogram corresponds to the average mean squared error (MSE) predicted by the indirect supervised learning model, and the error bar corresponds to the standard deviation of the MSE of 10 repeated experiments. It can be seen that, as the batch size increases, the model prediction accuracy increases. Secondly, the model performances when the training data length is 1, 20, 50, 100, 150, and 200 are evaluated, as shown in Figure 4b. The final result reveals that the best training data length is 150 for this problem. It is worth mentioning that the model performs poorly when the training data length is 1, mainly because the LSTM neural network with a sequence length of 1 is equivalent to a fully connected neural network. This result also reflects the advantages of LSTM over conventional neural networks, which is consistent with existing literature[12]. Finally, the effect of the batch normalization layer is examined, as shown in Figure 4c. Therefore, in subsequent experiments, the batch size and the training length are set to be 128 and 150, respectively, and the model uses a batch normalization layer.

In order to show the performance of the indirect supervised learning more intuitively, the prediction result of one randomly extracted well is plotted, as shown in Figure 5. The blue curve is the ground truth of the UCS, the orange curve represents the model prediction, and the red boxes show the enlarged details. One can see that the model based on the physics-constrained indirect label and the physical mechanism can accurately predict the trend of the UCS log, and the prediction result is close to the ground truth. In well log interpretation, experts mainly refer to the trend of well logs, and thus the prediction result of this model meets the needs in practice.

## IV. CONCLUSION AND DISCUSSION

This study uses the physical mechanism between variables as *a priori* information for supervised learning. The information about the variable to be predicted is required in conventional supervised learning. However, we proposed that this information can come not only directly from the variable (label), but also from the physical mechanism and other related variables (indirect labels). Therefore, we attempt to describe the variable to be predicted and perform supervised learning through some other variables, and the mapping relationship between these variables and the variable to be predicted.

In order to achieve this goal, we split the mapping relationship between the indirect label and label into linear and non-linear parts, and construct a projection matrix to project the prediction result of the model into the range of values that conform to the physical mechanism. Then, this study proposes an indirect physics-constrained loss, which does not need to obtain the variable to be predicted. In addition, the research utilizes projection matrix normalization and prediction covariance analysis to ensure that the model can be fully trained. Finally, through iterative optimization of the indirect physics-constrained loss, the model can ensure that the prediction and the indirect labels conform to a given mapping relationship (physical mechanism), and accurately predict the unknown

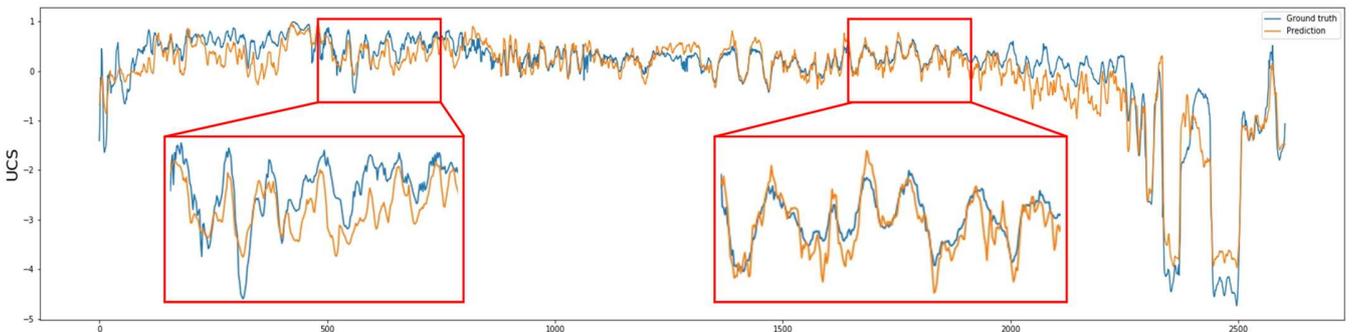
**Fig. 5.** The model prediction and ground truth of the UCS of the first test well in the last set of the experiments.

label.

In the experiment section, the feasibility of the physics-constrained indirect supervised learning is verified based on the well logs of 39 wells. We successfully predicted the UCS, which is not included in the training data, by using sonic logs as indirect labels and introducing the physical mechanism as *a priori* information into the loss function. This study shows that a loss function can be constructed based on the physical mechanism between the indirect label and the variable to be predicted, and it is possible to train a model by supervised learning based on indirect labels when the training data do not contain the variable to be predicted.

It should be mentioned that a naïve method can also be directly used to firstly learn and predict the indirect label (sonic logs) through a neural network, and then calculate the desired variable (UCS) based on the physical mechanism. This naïve method is similar to the physics-constrained indirect supervised learning and does not require the UCS as the training data, but differences still exist between these two methods.

Firstly, the physics-constrained indirect supervised learning provides a new modeling idea. When training neural networks, prior information is also used as a form of supervision in the loss function, and thus the supervised variable of the model is not the same as the desired variable. However, in the naïve method, the label is still required to train the model, which means that the supervised variable is consistent with the desired variable. Therefore, the two methods differ at the methodological level.

Secondly, the physics-constrained indirect supervised learning has the potential to further reduce the need for model constraint information, such as: (1) when some of the hyperparameters in the physical mechanism are unknown, it is not possible to calculate the desired variable by the naïve method. However, since the projection matrix in the physics-constrained indirect supervised learning does not depend on these hyperparameters, it is still theoretically possible to predict the desired variable; (2) because the physical mechanism is used to directly constrain the model's value space, the proposed model may have faster convergence speed and lower demand for training data; and (3) because the physical mechanism is integrated in the loss function, the same mechanism can be applied to different data sets, and thus the proposed model may offer superior generality. The above potential advantages and applications of the physics-constrained indirect supervised learning will be further studied in the future.

ACKNOWLEDGMENT

This work is partially funded by the National Natural Science Foundation of China (Grant no. 51520105005 and U1663208). The authors are grateful to Mr. Yuanqi Cheng for his assistance with data preparation, figure generation, and constructive discussions during the course of this work.